\documentclass{aa}  
\usepackage{graphicx}
\usepackage{color}
\usepackage{txfonts}
\begin{document} 

\title{Impact of Hot Inner Crust on Compact Stars at Finite Temperature}
% \titlerunning{}

\author{Clara Dehman,
\inst{1}\fnmsep\inst{2}\fnmsep\inst{3}\fnmsep\inst{4}\thanks{Email: clara.dehman@ua.es} Mario Centelles,\inst{5}\fnmsep\inst{6} \and Xavier Vi\~nas\inst{5}\fnmsep\inst{6}\fnmsep\inst{7}
          }

\institute{Departament de F\'isica Aplicada, Universitat d’Alacant, Ap. Correus 99, E-03080 Alacant, Spain 
\and
Nordita, KTH Royal Institute of Technology and Stockholm University, 10691 Stockholm, Sweden
\and 
Institute of Space Sciences (ICE-CSIC), Campus UAB, Carrer de Can Magrans s/n, 08193, Barcelona, Spain
   \and
Institut d'Estudis Espacials de Catalunya (IEEC), Carrer Gran Capità 2–4, 08034 Barcelona, Spain
\and 
Departament de Física Quàntica i Astrofísica (FQA), Universitat de Barcelona (UB), Martí i Franquès 1, 08028 Barcelona, Spain
\and 
Institut de Ciències del Cosmos (ICCUB), Universitat de Barcelona (UB), Martí i Franquès 1, 08028 Barcelona, Spain
\and 
Institut Menorquí d’Estudis, Camí des Castell 28, 07702 Maó, Spain}

% \date{Received 2024; accepted ,}
\date{}

\abstract{
We conducted a study on the thermal properties of stellar matter with the nuclear energy density functional BCPM. 
This functional is based on microscopic Brueckner-Hartree-Fock calculations and has demonstrated success in describing cold neutron stars. To enhance its applicability in astrophysics, in this study we extend the BCPM equation of state to finite temperature for $\beta$-stable neutrino-free matter, taking into consideration the hot inner crust. Such an equation of state holds significant importance for hot compact objects, particularly those resulting from a binary neutron star merger event. Our exploration has shown that with increasing temperature there is a fast decrease of the crust-core transition density,
suggesting that for hot stars it is not realistic to assume a fixed value of this density. The microscopic calculations also reveal that the presence of nuclear clusters persists up to $T=7.21$\,MeV, identified as the limiting temperature of the crust.
Above this threshold, the manifestation of clusters is not anticipated. Below this temperature, clusters within the inner crust are surrounded by uniform matter with varying densities, allowing for the distinction between the upper and lower transition density branches. Moreover, we computed mass--radius relations of neutron stars, assuming an isothermal profile for $\beta$-stable neutron star matter at various temperature values. Our findings highlight the significant influence of the hot inner crust on the mass--radius relationship, leading to the formation of larger and more inflated neutron stars. Consequently, under our prescription, the final outcome is a unified equation of state at finite temperature.
}


\maketitle

\section{Introduction}
Extensive efforts have been directed towards the dynamical simulations of core-collapse supernovae explosions \citep{pons1999,liebendoerfer2005,oertel2017,burgio2018} and the subsequent formation and evolution of a proto-neutron star (PNS). A realistic equation of state (EOS) of asymmetric nuclear matter over a wide range of densities and temperatures is one of the most vital inputs to these calculations \citep{bethe1990}. Current insights suggest three distinct phases in this process \citep{pons1999,barrere2022}:
(i) Roughly one second after the supernova core bounce, a comparatively cool central region, bordered by a hotter mantle, rapidly collapses, emitting neutrinos while accreting material.
(ii) Over the next 20 seconds or so, a slowly developing state of the PNS can be identified. The system first deleptonizes and heats up the interior parts of the star, and then begins to cool down further through neutrino diffusion.
(iii) After several minutes, the final state of the neutron star (NS) emerges, initially cooling by neutrino emission and later by photon emission from its surface \citep{pons2019}.

A cold NS becomes stratified in three primary regions: the outer crust, the inner crust, and a uniform core.
Each region is distinguished by its unique set of physical properties.
In the NS core, matter forms a homogeneous liquid comprising primarily neutrons, augmented by specific fractions of protons, electrons, and muons, ensuring the system remains in charge and $\beta$-equilibrium. At even greater densities deeper inside the core, hyperons, other strange particles, and deconfined quarks can emerge \citep{shapiro1983,haensel2007}. Transitioning from the core towards the outer layers, the density decreases, causing positive charges to cluster individually, identified by charge $Z$, and arrange into a solid lattice. This organization minimizes Coulomb repulsion between them. The lattice is embedded in a gas of neutrons and a background of electrons, ensuring overall charge neutrality for the system. This region is known as the inner crust. In the deepest layers of the inner crust, nuclear structures may take on non-spherical forms, commonly referred to as ``nuclear pasta'', driven by energy minimization \citep{baym1971a,lorenz1993}. Above the inner crust, in regions of lower densities, all neutrons become confined within nuclear clusters, giving rise to a lattice structure enriched with nuclei interspersed with a degenerate electron gas. This layer is defined as the outer crust \citep{baym1971b} and it extends from the interior, characterized by the neutron drip density, to the exterior, encompassing the envelope layer, which plays an important role in NS cooling \citep{potekhin2015,dehman2023}.
The crust of the NS, although it comprises a small fraction of its total mass and radius, is decisive in various observable signals from NSs, such as, for instance, glitches in pulsar NSs \citep{piekarewicz2014}, bursts and outbursts in intensely magnetized NSs \citep{beloborodov16,cotizelati18,dehman2020} and NS asteroseismology phenomena \citep{steiner2009,sotani2012,neill2023}.

The first observation of a NS merger event, GW170817, by \citet{ligovirgo2017}, has opened new avenues for studying the properties of matter under extreme conditions of densities and temperatures \citep{baiotti2019}. This event, resulting from the collision of two NSs, creates conditions that can lead to either the formation of a black hole or a notably massive NS. However, the final outcome of the GW170817 merger is still a subject of ongoing debate \citep{pooley2018}.
The pre-merger phase, known as the inspiral phase, leaves a unique imprint on the gravitational wave (GW) signal due to the tidal deformation of the stars involved, which is influenced by the EOS at zero or near zero temperature. This effect becomes noticeable early in the binary dynamics \citep{flanagan2008,hinderer2010,shibata2015,krastev2019}. The post-merger phase is expected to significantly heat the remnant, raising its temperature to tens of MeV. This heating affects post-merger characteristics, such as the remnant's lifespan, the GW spectrum, and the ejected mass, all of which are crucially dependent on the finite-temperature EOS \citep{oechslin2007,sekiguchi2011,bauswein2013,soma2020}.

A unified EOS for the crust and the core of NSs at zero temperature was introduced in \cite{sharma2015}, based on the BCPM nuclear energy density functional \citep{baldo2008,baldo2010,baldo2013,baldo2017}. In the BCPM functional, the bulk part of the energy density is entirely given by ab-initio Brueckner-Hartree-Fock (BHF) calculations of nuclear matter.
The bulk term is supplemented with surface and Coulomb terms, allowing one to describe the inhomogeneous nuclear structures of the NS crust and the homogeneous matter of the NS core simultaneously in a microscopic-based and consistent approach.
Despite significant progress in developing the EOS for cold dense matter, the EOS at finite temperature has not been as thoroughly investigated. 
Our study aims to introduce a suitable method for examining thermal effects on the BCPM EOS. We will particularly focus on the critical role of the hot inner crust.

Together with the growing interest in the high-energy astrophysical phenomena associated with compact stars, in recent years there has been much progress in the formulation of hot microscopic EOSs for the core of these systems (see, e.g., \cite{lu2019,logoteta2021,figura2021}, and references quoted therein).
Few calculations of hot EOSs are available, however, that include the crust at finite temperature computed with the same model as the core, and these models generally are phenomenological ones (see \cite{oertel2017,raithel2021} for recent reviews of hot EOSs).
Hence, in the present work we address a first exploration of a thermal EOS including the hot inner crust calculated on a microscopic basis by using the BCPM functional.
We develop the study within a neutrino-free $\beta$-stable framework, assuming an isothermal profile of hot nuclear systems consisting of neutrons ($n$), protons ($p$), electrons ($e$), and muons ($\mu$). 
In this context, the hot inner crust is present at relatively low densities and moderate temperatures. The mean free path of neutrinos in this regime is expected to be large enough for neutrinos not to be trapped \citep{reddy1999}.
For a broader scope, though, calculations with trapped neutrinos as well as configurations at constant entropy per baryon should also be addressed.
The present thermal EOS is particularly relevant for compact objects in warm environments where neutrinos are not trapped and matter is $\beta$-equilibrated.
This includes scenarios such as the post-merger phase of binary NS events or the last stages of the evolution of PNSs \citep{oertel2017,kumar2020,bethe1990}, assuming that the temporal evolution is slow enough to justify that matter approaches the indicated conditions \citep{lu2019}.

This paper is organised as follows. \S\ref{sec: equation of state of hot stellar matter} describes the BCPM EOS. \S\ref{subsec: zero temp} and \S\ref{subsec: finite temp} detail the EOS at zero and finite temperatures, respectively. The results obtained in this study are illustrated in \S\ref{sec: results and discussion}, where we highlight the significant influence of the hot inner crust. The conclusions are drawn in \S\ref{sec: conclusion}.

%%%%%%%%%%%%%%%%%%%%%%%%%%%%%%%%%%%%%%%%%%%%%%%%%%%%%%%%%%%%%%%
% BCPM and EOS of a beta-stable matter
%%%%%%%%%%%%%%%%%%%%%%%%%%%%%%%%%%%%%%%%%%%%%%%%%%%%%%%%%%%%%%%
\section{BCPM and Equation of State of a $\beta$ stable matter}
\label{sec: equation of state of hot stellar matter}

\subsection{Zero-Temperature Equation of State}
\label{subsec: zero temp}

The microscopic BHF calculations can be directly employed to obtain the EOS of the liquid core of NSs, where the nuclei have dissolved into their constituent protons and neutrons. However, a BHF calculation of finite nuclei and nuclear structures in a NS crust is not yet feasible. To describe finite nuclei, retaining as much information from the ab-initio BHF calculations as possible, the BCPM nuclear energy density functional was developed \citep{baldo2008, baldo2010,baldo2013,baldo2017}. Initially designed for characterizing the ground state of finite nuclei, the BCPM functional comprises a bulk component given by the BHF results in symmetric nuclear matter and neutron matter through the local density approximation, supplemented with a finite-range term to account for the surface properties. The Coulomb, spin-orbit, and pairing contributions are also included \citep{baldo2008, baldo2013}.
Having in total four adjustable parameters (for the surface and spin-orbit terms only), BCPM describes the
properties of finite nuclei with the same success as usual nuclear functionals that contain many more free parameters.
The bulk component of the model was obtained using the BHF approach, where the calculations employed the Argonne $v_{18}$ nucleon-nucleon (NN) interaction,
with the inclusion of three-body forces reduced to a two-body density-dependent term (\cite{wiringa1995}, chapter 1 of \cite{bookbaldo1999} and references therein, \cite{taranto2013}). The resulting EOS for both symmetric and asymmetric nuclear matter satisfies several criteria set by heavy ion collisions and recent   astrophysical observations.

In the BCPM functional, the bulk contribution to the energy per particle is the sum of the kinetic energy per particle
($e_\text{kin}$) and the potential energy per particle ($v_\text{int}$):
\begin{equation}
    e(n,\beta)=e_{\text{kin}}(n,\beta)+v_{\text{int}}(n,\beta).
    \label{total energy T0}
\end{equation}
The kinetic energy per particle at zero temperature is that of a non-interacting cold Fermi gas with degeneracy factor of $2$ and isospin asymmetry $\beta$:
\begin{equation}
    e_{\text{kin}}(n,\beta)= \frac{1}{2} \frac{3}{5}\frac{\hbar^2}{2m}\bigg(\frac{3 \pi^2n}{2} \bigg)^{2/3} \bigg[ \big( 1 + \beta\big)^{5/3} 
+  \big( 1 - \beta\big)^{5/3}\bigg],
\end{equation}
where $n=n_n+n_p$ ($n_n$ and $n_p$ are, respectively, the neutron and proton number densities) and $\beta=(n_n-n_p)/n$ is the isospin asymmetry parameter. The potential energy per particle in the bulk, $v_\text{int}(n,\beta)$, is represented as a quadratic interpolation between the interaction energy per particle in symmetric nuclear matter (SNM), $v_\text{int}(n,0)$, and pure neutron matter (PNM), $v_\text{int}(n,1)$:
\begin{equation}
  v_{\text{int}}(n,\beta)=  v_{\text{int}}(n,0)+  \bigg(v_{\text{int}}(n,1)- v_{\text{int}}(n,0)\bigg)~\beta^2.
  \label{eq: Vint}
\end{equation}
Here, $v_\text{int}(n,0)$ and $v_\text{int}(n,1)$ are given by the microscopic BHF calculations at various densities.
For computational efficiency, in BCPM an accurate polynomial fit of the discrete BHF points is performed as a function of the density $n$ \citep{baldo2010,sharma2015}. This fit remains valid up to a density \mbox{$n \approx 0.4$} fm$^{-3}$. For bulk matter at higher density values, as the ones that can be found in NS cores, we utilize functional forms that provide excellent parametrizations of the BHF results for SNM and PNM at $T=0$ \citep{burgio2010,sharma2015}.

The BCPM functional was employed in \cite{sharma2015} to establish a unified EOS for the outer crust, inner crust and  core of NSs at zero temperature. Many-body calculations of the inhomogeneous structures in the NS crust currently fall beyond the scope of the BHF approach, which is used for modeling the homogeneous core. The calculation of the NS crust with the BCPM functional maintains a consistent microscopic approach in describing the entire stellar structure. The NS crust is modeled within the Wigner-Seitz (WS) cell approximation, dividing space into non-interacting cells, each containing a single nuclear cluster in charge and $\beta$-equilibrium. In the outer crust, matter is composed of fully ionized atomic nuclei, forming a solid lattice to minimize Coulomb repulsion, permeated by a degenerate electron gas. The critical data for constructing the outer crust EOS are the nuclear masses, sourced from the AME2012 evaluation \citep{audi2012}, or, if they are unknown, calculated using the Hartree-Fock-Bogoliubov (HFB) method with the BCPM functional \citep{sharma2015}. As the star's average density increases, nuclei become increasingly neutron-rich, eventually leading to the inner crust, where neutrons begin to drip. The inner crust thus consists of nuclear clusters immersed in a gas of dripped neutrons and a background of electrons.
In the bottom of the inner crust, surrounding the uniform core, a thin layer of pasta phases (nuclear clusters that adopt non-spherical shapes) is predicted by the BCPM calculations \citep{sharma2015}.

A fully quantum calculation of the inner crust is a challenging task due to the presence of the neutron gas. To describe the cold inner crust in \cite{sharma2015}, self-consistent Thomas-Fermi (TF) calculations with BCPM were performed. This approach offers a significant advantage, because the EOS in the inner crust is primarily influenced by the neutron gas, which means that shell and pairing effects have a marginal impact on the EOS
and the proton fraction \citep{pearson2022}. Using this formalism, the EOS of the crust and the core are both determined through the BCPM energy density functional. The generalisation of this formalism to finite temperature for the core and the inner crust is presented in
\S\ref{subsec: finite temp}.

\subsection{Equation of State at Finite Temperature}
\label{subsec: finite temp}

In this work we extend the BCPM EOS to finite temperature. We introduce the thermal effects through the Fermi occupation numbers in the energy density functional, while keeping the interactions the same as at $T=0$, as usually done in nuclear functionals. 
This approach is often known as the frozen-correlations approximation \citep{baldo1999,burgio2007}, as it assumes the nucleon single-particle potential at finite $T$ to be the same as at $T=0$. It has been checked in hot BHF calculations that this assumption is verified with good accuracy, at least for temperatures up to about 30 MeV \citep{baldo1999,burgio2007,burgio2010,lu2019}.

\subsubsection{The Liquid Core}
\label{sec: liquid core}

To determine the self-consistent occupation numbers in asymmetric nuclear matter at a given temperature, we minimize the thermodynamic potential, given by:
\begin{eqnarray}
    \Omega &=&  \sum_{q=n,p} \sum_{k} E_q (n_n(k),n_p(k)) - TS - \sum_{q=n,p} \sum_{k} \mu_q n_q(k),  \nonumber \\ 
&=& \sum_{q=n,p} \sum_{k} F_q (n_n(k),n_p(k)) - \sum_{q=n,p} \sum_{k} \mu_q n_q(k) .
    \label{eq: grand potential}
\end{eqnarray}
Here, $E$ represents the internal energy, $T$ is the temperature, $F$ is the free energy, $S$ is the entropy, and $\mu_q$ and $n_{q}(k)$ are, respectively, the chemical potential and the occupation number of each type of nucleon, with $q=n,p$.
The entropy per particle for asymmetric nuclear matter with proton and neutron fraction $y_q=n_q/n$ at finite temperature is expressed as:
\begin{equation}
s(n,\beta,T)= \sum_{q=n,p} y_q s_q(n,\beta,T)  ,
    \label{eq: entropy asym T}
\end{equation}
with $s_q$ being the entropy per particle of each component:
\begin{equation}
    s_q(n,\beta,T) = - \sum_{k} \bigg( n_q(k) \ln n_q(k) + \big[1 - n_q(k)\big] \ln \big[1-n_q(k) \big] \bigg). 
    \label{eq: entropy for p or n asym T}
\end{equation}
In eqs.\,\eqref{eq: grand potential} and \eqref{eq: entropy for p or n asym T}, $n_q(k)$ represent the Fermi occupation numbers, which are the solutions of the variational equations obtained by applying the variational principle to the grand potential (eq.\,\eqref{eq: grand potential}). These occupation numbers read as:
\begin{equation}
    n_q(k)= \frac{1}{1+e^{(\varepsilon_q(k)-\mu_q)\hspace{0.5mm}/\hspace{0.5mm} T}}, 
    \end{equation}
where $\varepsilon_q(k)$ denotes the single-particle spectrum given by:
\begin{equation}
   \varepsilon_q(k) =\frac{\hbar^2 k^2}{2m}+V_q(n,\beta).
\end{equation}
In the frozen-correlations approximation, the single-particle potential $V_q$ depends on the baryon number density $n$ and the isospin asymmetry $\beta$ and is independent of temperature. It is given by:
\begin{eqnarray}
       V_q(n,\beta) = \frac{\partial \big(n \, v_\text{int}(n,\beta)\big)}{\partial n_q}
  = v_\text{int}(n,\beta) + n \frac{\partial v_\text{int}(n,\beta)}{\partial n_q}~,
    \label{eq: Vq}
\end{eqnarray}
with $v_\text{int}(n,\beta)$ of the cold calculation, eq.\,\eqref{eq: Vint}.
The number density $n_{q}$ of neutrons and protons reads as:
\begin{equation}
    n_q= \frac{2}{(2 \pi)^3}\int_0^\infty n_q(k)\hspace{0.5mm} d^3 k =
\frac{1}{2\pi^2}\Bigg(\frac{2mT}{\hbar^2}  \Bigg)^{3/2} J_{1/2}(\eta_q) ,
\label{eq: density uniform matter}
\end{equation}
where $J_{1/2}(\eta_q)$ is the Fermi integral of index $\nu=1/2$:
\begin{equation}
  J_{1/2}(\eta_q)= \int_0^\infty \frac{z^{1/2}dz}{1+ e^{(z-\eta_q)}}. 
  \label{eq: J1/2}
\end{equation}
Here, we have $z=[ \varepsilon_q(k)- V_q ]\hspace{0.5mm}/\hspace{0.5mm}T$ and the fugacity reads $\eta_q=[\mu_q - V_q]\hspace{0.5mm} / 
\hspace{0.5mm}T$.

At finite temperature, the total energy per particle of asymmetric nuclear matter becomes
\begin{equation}
    e(n,\beta,T)=e_{\text{kin}}(n,\beta,T)+v_{\text{int}}(n,\beta),
    \label{eq: energy per particle}
\end{equation}
where the interaction term, $v_\text{int}(n,\beta)$, is defined in eq.\,\eqref{eq: Vint},
and the nucleonic kinetic energy per particle reads as:
\begin{equation}
\frac{ 1 }{n} \sum_{q=n,p}  \frac{\hbar^2 \, \tau_q}{2m}, 
    \label{eq: kinetic energy per particle}
\end{equation}
where
\begin{equation}
\tau_q= \frac{2}{(2 \pi)^3}\int_0^\infty k^2 \, n_q(k) \hspace{0.5mm} d^3 k =
\frac{1}{2 \pi^2}\Bigg( \frac{2mT}{\hbar^2}  \Bigg)^{5/2} J_{3/2}(\eta_q) ,
\end{equation}
and $J_{3/2}(\eta_q)$ is the Fermi integral of index $\nu=3/2$:
\begin{equation}
    J_{3/2}(\eta_q)= \int_0^\infty \frac{z^{3/2} dz}{1+e^{(z-\eta_q)}}.
\end{equation}
The proton and neutron chemical potentials are expressed in terms of the temperature $T$, fugacity $\eta_q$ (obtained by inverting eq.~\eqref{eq: density uniform matter} for a given nucleon density $n_q$), and the single-particle potential $V_{q}$:
\begin{equation}
\mu_q(n,\beta,T)= \eta_q T + V_{q}(n,\beta) .
    \label{eq: chem pot asym finite T}
\end{equation}
The pressure is given by:
\begin{equation}
P\,(n,\beta,T)=n \, \bigg[\sum_{q=n,p}y_q \, \mu_q(n,\beta,T) -f(n,\beta,T)  \bigg] ,
\end{equation}
where $f(n,\beta,T)$ is the nucleonic free energy per particle.

In homogeneous neutrino-free stellar matter containing nucleons and leptons (electrons and muons), under the constraint of charge neutrality ($n_p = n_e + n_\mu$), the conditions for $\beta$-equilibrium are:
\begin{equation}
    \mu_n - \mu_p = \mu_e\,, \quad \mu_{\mu} = \mu_e.
\end{equation}
In our approach, nucleons are treated as non-relativistic particles, while the leptons are considered as relativistic free particles. With the inclusion of leptons, the energy in eq.\,\eqref{eq: energy per particle} also contains the leptonic contribution:
\begin{equation}
 \frac{1}{n} \frac{2}{(2 \pi)^3} \int_0^{\infty}    \varepsilon_l(k)  \, n_{l}(k) \, d^3k  , \quad\quad l=e,\mu,
 \label{eq: ekin leptons}
\end{equation}
where $\varepsilon_l(k)= \sqrt{\hbar^2 k^2 c^2+ m_{l}^2 c^4}$
and $n_{l}(k)$ is the occupation number of leptons:
\begin{equation}
     n_{l}(k) =   \frac{1}{1+e^{(\varepsilon_l(k)-\mu_l)\hspace{0.5mm}/\hspace{0.5mm} T}} ~. 
\end{equation}
Likewise, the total entropy and the total pressure of the system are obtained by adding the contribution of free leptons to that of nucleons. 

%%%%%%%%%%%%%%%
% Inner crust
%%%%%%%%%%%%%%%
\subsubsection{The Hot Inner Crust}
\label{sec: inner crust}

We have employed the TF method at finite temperature to calculate the EOS of the inhomogeneous matter of the crust using the BCPM functional. The nuclear clusters in the hot crust are computed inside spherical WS cells of radius $R_c$. Each cell is electrically neutral, and interactions between cells are neglected. At nuclear densities, the electrons exhibit highly relativistic behavior, as their Fermi momenta far exceed their rest mass. Therefore, they can be assumed to be uniformly distributed within the WS cell. Inside the cell, we impose $\beta$-equilibrium, leading to the condition $\mu_n = \mu_p + \mu_e$ when neutrinos have left the star. Similarly to the core, the properties of the hot system within the WS cell are determined by minimizing the thermodynamic potential (eq.\,\eqref{eq: grand potential}), which, in the crust, also includes non-uniform contributions (see below). It is important to note that at finite $T$, nuclei become unstable against nucleon evaporation, giving rise to a surrounding gas of evaporated nucleons. Thus, within the WS cell, a coexistence between the nuclear cluster plus gas and the gas alone occurs. Consequently, in the hot TF calculations in a WS cell there exist two solutions of the TF equations in equilibrium: one corresponding to the liquid-plus-gas (LG) phase and the other one corresponding to the gas (G) phase alone (\cite{suraud1987,sil2002}; and references therein).

In the context of the crust, the internal energy contributing to the thermodynamic potential is expressed as follows within the WS cell of volume $V_c$:
\begin{equation}
E = \int_{V_c} \bigg[\mathcal{H}(n_n, n_p) + \mathcal{E}_{el} + \mathcal{E}_\text{coul} + \mathcal{E}_{ex} + m_pn_p + m_n n_n\bigg] \,d\boldsymbol{r}.
    \label{eq: total energy crust}
\end{equation}
Here, $\mathcal{H}(n_n, n_p)$ represents the nuclear energy density, where $n_n =n_n\boldsymbol{r})$ and $n_p=n_p(\boldsymbol{r})$ are the neutron and proton number densities, respectively, which are position dependent within the cell. In the hot TF approach, $\mathcal{H}$ incorporates the kinetic energy density of protons and neutrons at finite temperature (see eq.\,\eqref{eq: kinetic energy per particle}). Additionally, it incorporates the cold interacting part $\mathcal{V}(n_n,n_p)$, as determined by the BCPM functional, covering both bulk (eq.\,\eqref{eq: Vint}) and surface contributions \citep{sharma2015}.
The surface term is taken as at zero temperature:
\begin{eqnarray}
\mathcal{E}_{\tt surf}(n_n, n_p) &=& \frac{1}{2} \sum_{q,q'} n_q(\boldsymbol{r})   \int  v_{q q'}(\boldsymbol{r}-\boldsymbol{r}')
n_{q'}(\boldsymbol{r}')  d\boldsymbol{r}' \nonumber\\
&& \mbox{} - \frac{1}{2}\sum_{q,q'}  n_q(\boldsymbol{r})  n_{q'}(\boldsymbol{r})   \int v_{q q'} (\boldsymbol{r}') d\boldsymbol{r}'. 
    \label{eq: surface energy density}
\end{eqnarray}
The second term in eq.\,\eqref{eq: surface energy density} is subtracted to avoid contamination of the the bulk part derived from the microscopic nuclear matter calculations. For the finite-range form factors, we use the same Gaussian shape $v_{q q'}(r)$ as in \cite{baldo2010,sharma2015}. 
The term $\mathcal{E}_{el}$ in eq.\,\eqref{eq: total energy crust} represents the energy density arising from the motion of electrons at finite temperature, with their energy per particle given in eq.\,\eqref{eq: ekin leptons}. For the densities and temperatures of our calculations of the crust, muons did not occur in this region of the star. 

At finite temperature, the direct Coulomb contribution in a WS cell corresponding to the LG and G phases, as discussed in \cite{sil2002}, is expressed as:
\begin{equation}
\mathcal{E}^\text{coul}_{LG}(n^p_{LG}, n_e) = \frac{1}{2}  \big(n^p_{LG}(\boldsymbol{r}) - n_e \big) 
\big( V^\text{coul,$p$}_{LG}(\boldsymbol{r}) - V_e(\boldsymbol{r}) \big),
 \label{eq: Ecoul cold inner crust liquid-gas}
\end{equation}
and
\begin{eqnarray}
\mathcal{E}^\text{coul}_{G}(n^p_{G}, n_e) &=& \frac{1}{2}  \big(n^p_{G}(\boldsymbol{r}) - n_e \big) 
\big( V^\text{coul,$p$}_{G}(\boldsymbol{r}) - V_e(\boldsymbol{r}) \big) 
\nonumber \\ 
&& + \, n^p_{L}(\boldsymbol{r})
\big( V^\text{coul,$p$}_{G}(\boldsymbol{r}) - V_e(\boldsymbol{r}) \big),
\label{eq: Ecoul cold inner crust gas}
\end{eqnarray} 
where $n^p_{L}=n^p_{LG}-n^p_G$ is the proton density of the nuclear cluster.
The potentials $V^\text{coul,$p$}_{LG(G)}(\boldsymbol{r})$ and $V_e(\boldsymbol{r})$ are given by:
\begin{equation}
 V^\text{coul,$p$}_{LG(G)}(\boldsymbol{r}) = \int \frac{e^2 n^p_{LG(G)}(\boldsymbol{r}^{'})}{|\boldsymbol{r} - \boldsymbol{r}^{'} |}
d\boldsymbol{r}^{'}, \hspace{5mm}
V_e(\boldsymbol{r}) = \int \frac{e^2 n_e}{|\boldsymbol{r} - \boldsymbol{r}^{'} |} d\boldsymbol{r}^{'}.
 \label{eq: Vp-LG-G and Ve}
\end{equation}
The direct part of the single-particle Coulomb potential, obtained by performing the functional derivatives of the LG and G direct Coulomb energies with respect to $n_{LG}$ and $n_G$, respectively, is the same for both phases and reads:
\begin{equation}
 V^d_\text{coul}(\boldsymbol{r}) = \int \frac{e^2\big( n^p_{LG}(\boldsymbol{r}^{'})-n_e \big)}{|\boldsymbol{r} - \boldsymbol{r}^{'} |}
d\boldsymbol{r}^{'}.
 \label{eq: Vdir_c}
\end{equation}
The total Coulomb energy in the LG and G phases is comprised of the direct contributions, as described in eqs.\,\eqref{eq: Ecoul cold inner crust 
liquid-gas} and \eqref{eq: Ecoul cold inner crust gas}, respectively, along with the exchange contribution from protons and electrons, which is calculated at the Slater level:
\begin{equation}
\mathcal{E}^{ex}_{LG(G)}(n^p_{LG(G)}, n_e) =
- \frac{3}{4} \Bigg(\frac{3}{\pi} \Bigg)^{1/3} e^2 
\bigg( n^p_{LG(G)}(\boldsymbol{r})^{4/3} + n_e^{4/3} \bigg).
\label{eq: exchange energy inner crust}
\end{equation}

Taking functional derivatives of the thermodynamical potential (eq.\,\eqref{eq: grand potential}) with the energy given by eq.\,\eqref{eq: total energy crust}, which includes Coulomb effects in both LG and G phases, with respect to $n^q_{LG}$ and $n^q_{G}$ for neutrons and protons, we obtain the following set of coupled equations, as described in \cite{sil2002}:
\begin{equation}
T \eta^q_{LG}(\boldsymbol{r}) + V^q_{LG}(\boldsymbol{r})+ V^\text{coul}_{LG}(\boldsymbol{r})=\mu_q,
\end{equation}
\begin{equation}
T \eta^q_{G}(\boldsymbol{r}) + V^q_G(\boldsymbol{r}) + V^\text{coul}_G(\boldsymbol{r})=\mu_q.
\end{equation}
Here, $V^q_{LG(G)}$ represents the nuclear part of the single-particle potential in the LG(G) phases, and $V^\text{coul}_{LG(G)}$ is the total Coulomb potential, which is the sum of the direct term from eq.\,\eqref{eq: Vdir_c} and the exchange part obtained from derivatives with respect to $n^p_
{LG}$ ($n^p_{G}$) of the exchange energy as given in eq.\,\eqref{eq: exchange energy inner crust}.

For a certain temperature $T$ and average baryon density $n$ in the WS cell,
assuming a radius $R_c$ (equivalently, a baryon number $A$) for the cell,
the set of variational equations is solved self-consistently using the method described in \cite{sil2002}.
This method allows for the determination of the composition $(A,Z)$ of minimal free energy per baryon in $\beta$-equilibrium. Next, we search for the optimal cell size (i.e., optimal baryon number) for the given baryon density $n$ by repeating the calculation for various values of $R_c$.
In the present study at finite $T$, we restrict ourselves to spherical WS cells.
This choice is motivated by the fact that in the BCPM calculations at $T=0$, nuclear pasta shapes appear just in a small density region near the transition to the core, cf.\ Fig.\,4 of \cite{sharma2015}.
Furthermore, as the relevant energy surfaces are very flat near the optimal configurations and they present tiny energy differences between spherical and non-spherical cells,
the different geometrical shapes have a negligible impact on the EOS (pressure vs density) \citep{sharma2015,pearson2020}.
Therefore, for simplicity, here we consider spherical cells only. However, the consideration of pasta phases would be important in applications
where the detailed structure of the crust is required, as pasta may strongly influence elasticity, transport and other properties of the NS crust \citep{haensel2007}.

We note that at finite $T$, the distinctive nuclear shell effect is eroded, as thermal excitation disrupts the well-defined energy levels within the nucleus present at $T=0$. It is a well-known consequence of the smearing of the Fermi surface caused by the Fermi occupation numbers.
In hot nuclear systems with temperatures higher than $T\approx2$--3\,MeV the shell effects have vanished \citep{barranco1981,brack1985,pi1986}, meaning that these systems are optimal objects for TF calculations like ours.

We are focused on calculating the EOS in various regions of the NS, with a specific emphasis on determining the structure and EOS in the crust using the BCPM model at a finite $T$. To obtain the EOS in the inner crust, it is necessary to compute the pressure, which is determined by taking appropriate derivatives of the energy with respect to the size of the WS cell.
As shown in Appendix A of \cite{sharma2015} (also see \cite{pearson2018}), at $T=0$ the pressure in the inner crust is a result of the neutron and electron gases within which the nuclear structures are embedded.
At finite $T$, it consists of the contributions of the nucleons (neutrons and protons) and the free electrons in the gas phase, plus the Coulomb exchange pressure of the charged particles in the gas.

The exploration of the presence of hyperons and other exotic degrees of freedom in hot supranuclear matter is extremely interesting and a major topic of ongoing research \citep{oertel2017,marques2017}.
We recall, however, that since we are largely focused on the regime of densities of the crust, where hyperons are not expected to occur in any significant amount \citep{menezes2017}, we consider the picture where the baryonic composition of the star corresponds to purely nucleonic matter.
Recent Bayesian analyses of the possible behaviors of the EOS of dense matter indicate that the existing astrophysical measurements of NSs are compatible with the fully nucleonic hypothesis for the composition of dense matter \citep{thi2021}.

\section{Impact of Hot Crust}
\label{sec: results and discussion}

To study the EOS in the NS crust at finite temperature, we start by examining the transition between crust matter and uniform matter across different temperature ranges. Although it is possible to estimate the crust-core transition density from the core side by looking at the threshold for the instability of uniform matter against clustering \citep{kubis2004,xu2009,moustakidis2012,gonzalez2019}, in this study, we seek the transition density from the perspective of the crust to emphasize the behavior of the crust at finite temperature.
In the inner crust, we search for the nuclear composition that provides the minimal free energy per baryon in $\beta$-equilibrium using the hot TF method discussed in the previous section. This calculation of the crust includes self-consistently the Coulomb and surface effects, which are absent in uniform matter.
As in the $T=0$ case \citep{sharma2015}, the transition from the crust to the uniform matter is determined by an energy criterion. This transition occurs when the free energy per baryon of the homogeneous phase is lower than that obtained for the clustered phase within the WS cell. At $T=0$ the BCPM spherical-cell calculations predict that the crust-core interface is located at baryon density $n=0.08$\,fm$^{-3}$.

\begin{figure}[t]
    \centering
    \includegraphics[width=8.5cm, height=8cm]
    {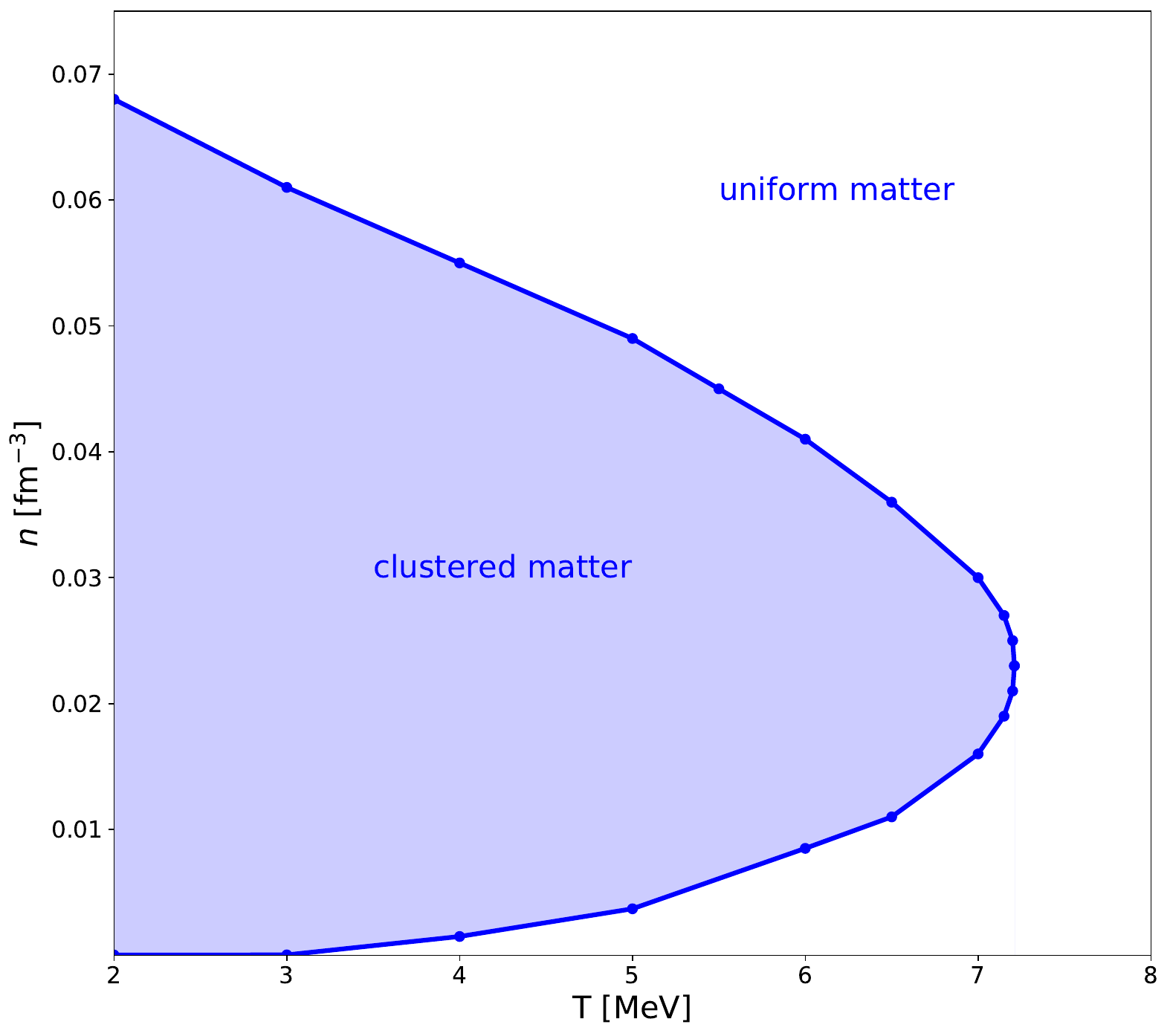}
    \caption{The two branches, upper and lower, of the transition density between the uniform matter and the clustered matter are computed at different temperature values. The blue shaded area corresponds to the presence of nuclear clusters, determined using TF calculations within a WS cell.}
    \label{fig: transition density}
\end{figure}

\begin{table}[ht]
    \centering
        \caption{Table showing the upper ($n_{t}^u$) and lower ($n_{t}^l$) transition densities, as illustrated in Fig.~\ref{fig: transition density}, for specific temperature values.}
    \begin{tabular}{|c|c|c|}
    \hline
  $T$ &  $n_{t}^u$& $n_{t}^l$ \\
   $[$MeV$]$ &  $[$fm$^{-3}]$& $[$fm$^{-3}]$ \\
  \hline
    \hline
2.00 & 0.068 & $3\times 10^{-5}$\\
3.00 & 0.061  &  0.0004 \\
4.00 &  0.055  &  0.0015\\
5.00 & 0.049 & 0.0037 \\
6.00 & 0.041  &  0.009\\
6.50 & 0.036 &  0.011\\
7.00 & 0.030 & 0.016\\
7.15 & 0.027 & 0.019\\
7.20 & 0.025 & 0.021\\
7.21 & 0.023 & 0.023\\
\hline
    \end{tabular}
    \label{tab: transition density}
\end{table}

In Fig.~\ref{fig: transition density}, we show the regions where the clustered and the uniform matter are the most stable phases as a function of the temperature and the average baryon density, derived from hot TF calculations with BCPM. This figure features a thick blue line in the density-temperature plane that separates these two regions. The figure also illustrates that the crust-uniform matter transition boundary, as a function of temperature, is a bivalued function that decreases with increasing temperature until it converges to a single endpoint. This endpoint occurs at a specific limiting temperature ($T_\mathrm{lim}$), predicted to be of 7.21\,MeV,
beyond which crust-uniform matter transitions are not feasible. The upper and lower transition densities illustrated in Fig.~\ref{fig: transition density} are collected in Table~\ref{tab: transition density} for specific temperature values. 
The concept of the upper transition density is analogous to the transition density between the inner crust and the core of NSs at zero temperature, whereas the lower transition density does not have a $T=0$ counterpart. 

\begin{figure}
    \centering
{\includegraphics[height=0.27\textheight]{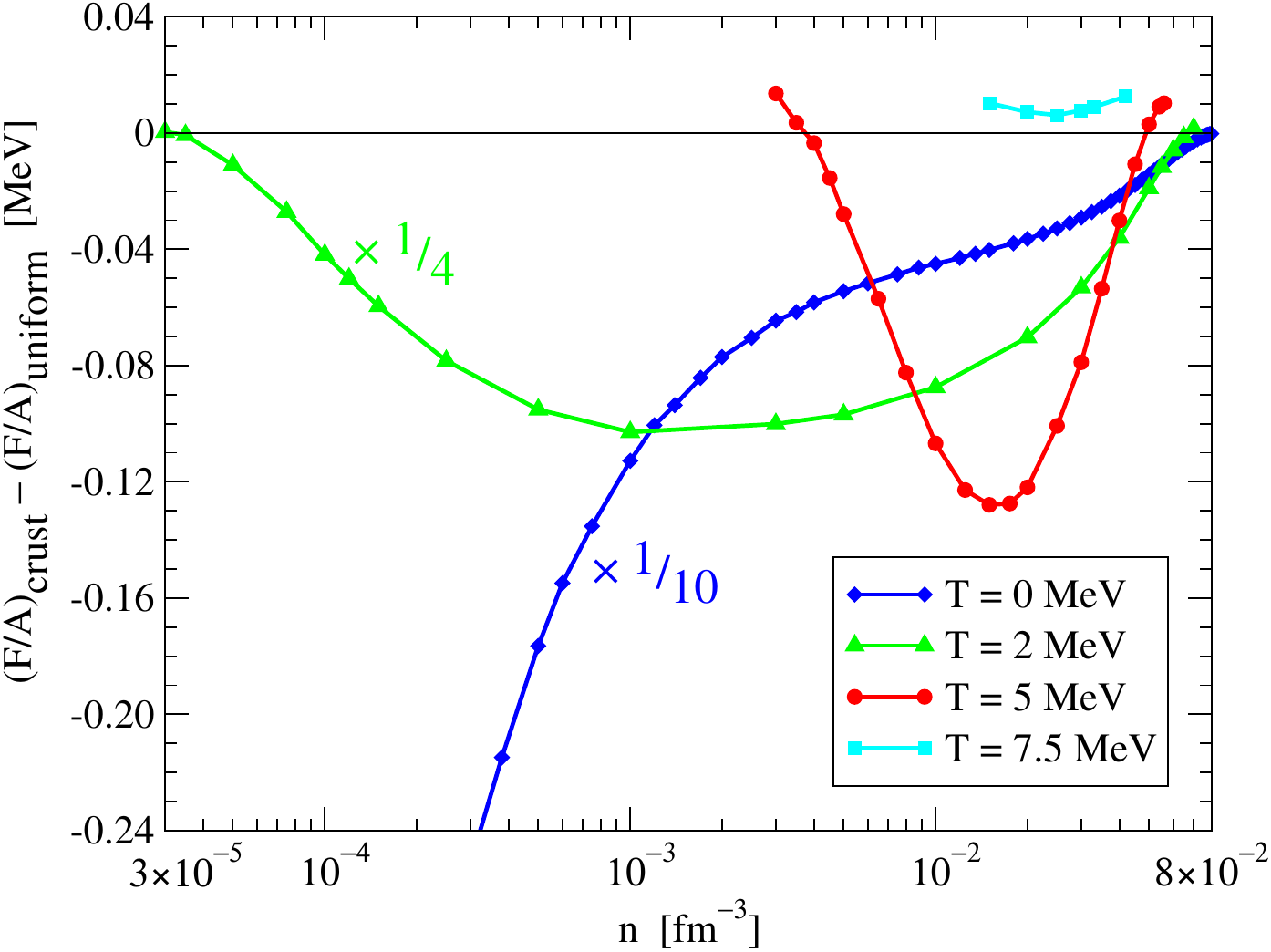}}
\caption{Difference between the calculated free energy per particle of the clustered matter and of the uniform matter for $T=0$, 2, 5, and 7.5\,MeV. Notice that the $T=0$ result is divided by a factor 10 and the $T=2$\,MeV result by a factor 4.}
\label{fig:FoAdiff}
\end{figure}

Fig.~\ref{fig:FoAdiff} displays the difference in free energy per baryon between crust matter and uniform matter for $T=0$, 2, 5, and 7.5\,MeV. 
At each temperature, the crust phase is favored while the result is negative. Starting from $T=0$, the gap between the free energy of the crust phase and the uniform phase is seen to shrink rapidly as $T$ increases. We note that in this figure we have divided the $T=0$ result by 10 and the $T=2$ result by 4 to display the four temperatures with the same vertical scale. It can also be observed that the $T=0$ curve crosses the zero axis of Fig.~\ref{fig:FoAdiff} only once, at $n=0.08$\,fm$^{-3}$, which is the crust-core transition density of the cold case. However, with raising $T$, the free energy gap between crust and uniform matter presents an inverted-bell shape,  
such that for $T<T_\mathrm{lim}$, as illustrated by the $T=2$ and $5$\,MeV cases, the gap vanishes at two different density points, corresponding to the lower ($n_t^l$) and the upper ($n_t^u$) transition density for the given temperature. The separation between $n_t^l$ and $n_t^u$ closes with higher $T$, until $n_t^l = n_t^u$ for the limiting temperature $T_\mathrm{lim}=7.21$\,MeV. Above $T_\mathrm{lim}$, the free energy of the clustered matter remains higher than for the uniform matter, as shown in the plot by the $T=7.5$\,MeV case.

Some differences are noticed between the zero and the finite temperature scenarios. At zero $T$, the inner crust of a NS consists of nuclear clusters that encapsulate all the protons present in the WS cell, amid a gas of dripped neutrons. As the average baryon density rises, the contribution from the neutron gas also increases. A threshold density exists, above which uniform matter becomes the most stable phase.
With thermal excitation, more neutrons evaporate from the clusters into the surrounding gas, resulting in an increased neutron number in the gas phase. Protons also evaporate at finite $T$, contributing to the formation of a charged gas. In this scenario, a lower transition density appears, below which uniform matter becomes again the most stable phase. Our calculations have shown that the reemergence of the uniform matter in the lower transition density branch is due to both the proton Coulomb contribution and the interaction between protons and neutrons in the gas phase.
As depicted in Fig.~\ref{fig: transition density} and seen also in Fig.~\ref{fig:FoAdiff}, the density range at a given $T$ in which clustered matter is the most stable phase diminishes with increasing $T$. The upper and lower boundaries defining this region converge, ultimately merging into a single point at the limiting temperature. This scenario is akin to the phenomena observed in nuclear matter when analyzing instabilities within a homogeneous medium \citep{barranco1981,lattimer1991,hempel2010}.

\begin{figure*}[t]
    \centering
{\includegraphics[height=0.25\textheight]{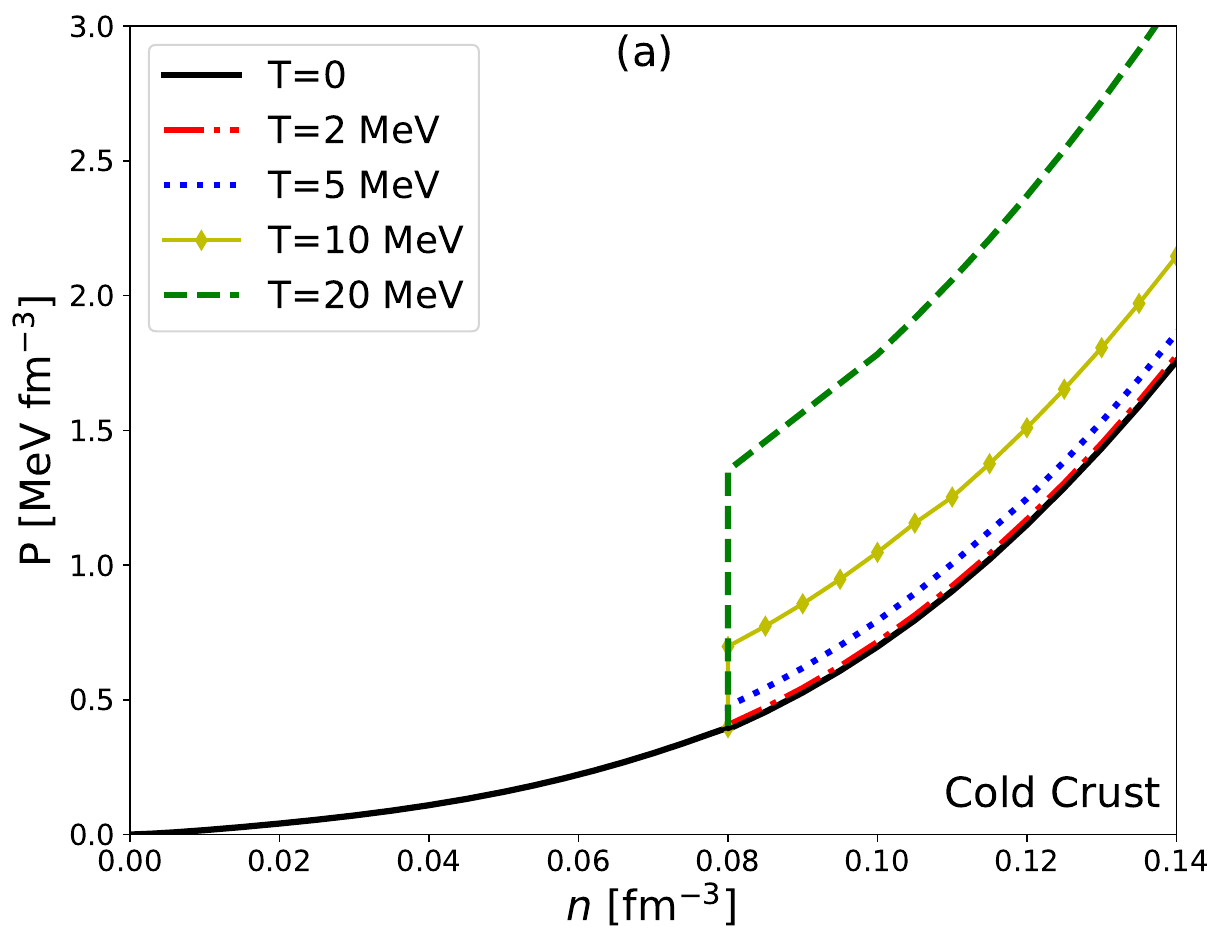}}
{\includegraphics[height=0.25\textheight]{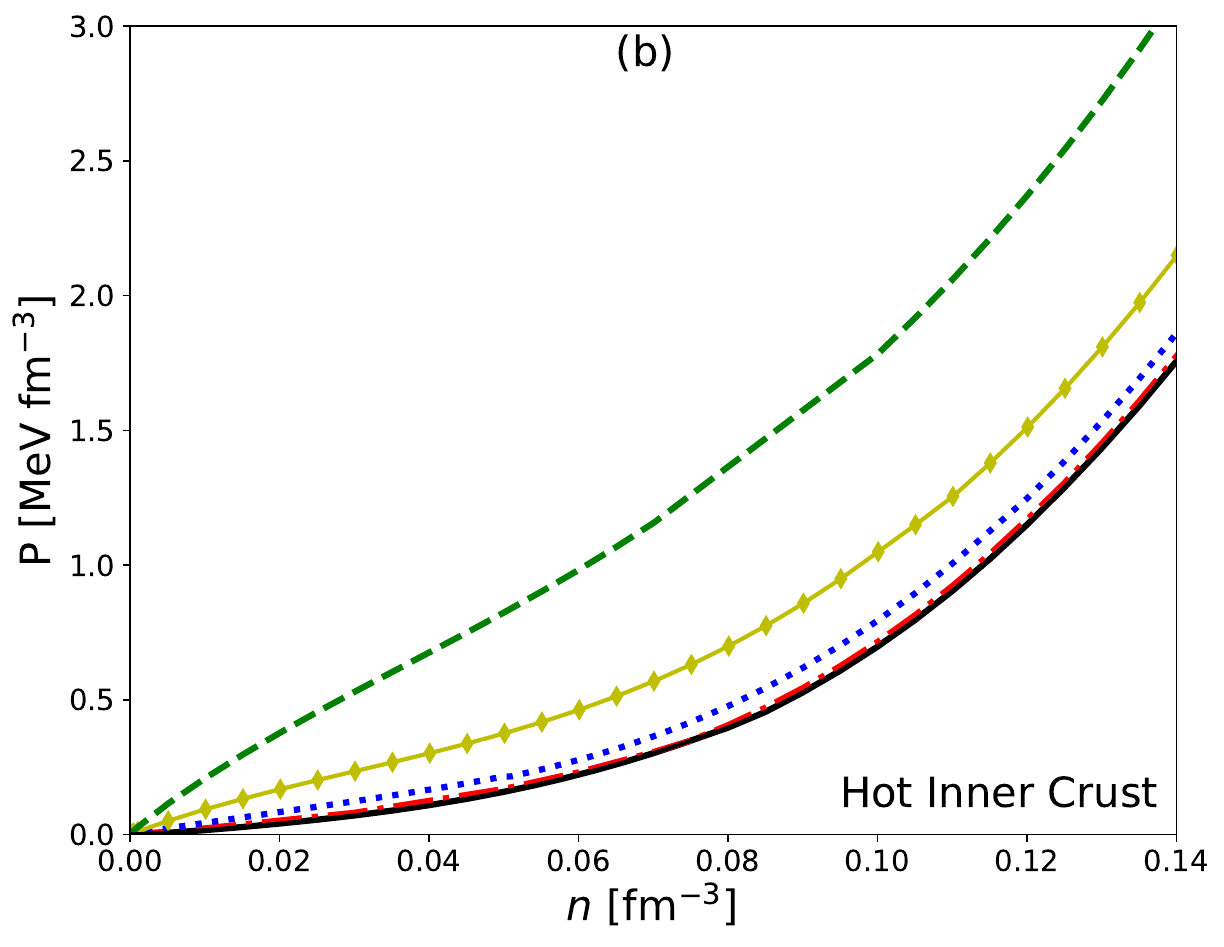}}
\caption{EOSs of $\beta$-stable neutrino-free matter at different temperature values of $T=0$, $2$, $5$, $10$, and $20$\,MeV. In the left panel, we use a cold crust assuming the transition between the cold crust and the uniform core fixed at $n=0.08$\,fm$^{-3}$. In the right panel, we use a hot inner crust
and the crust--uniform matter transition is determined as described in \S\ref{sec: results and discussion}, consistently with Fig.~\ref{fig: transition density}.}
\label{fig: EOS WITH COLD + hot CRUST}
\end{figure*}

It is noteworthy that at the transition density between the crust and uniform matter for both the lower and upper branches, our findings indicate that the proton fraction at these points is approximately \(0.03\). Additionally, we have computed the plasma parameter, defined as $\Gamma = (Z_\text{cl} e)^2 / (R_\text{c} T)$, at the transition density points illustrated in Table~\ref{tab: transition density} and Fig.~\ref{fig: transition density}. We find that $\Gamma$ is consistently less than 175 at all transition densities, indicating that the clusters are in the liquid phase \citep{pi1986}. This result is in agreement with previous findings reported in the literature \citep{aguilera2008}.

Certain studies of the NS EOS at finite temperature assume a cold crust because the EOS of the hot crust for the considered nuclear interaction is not available. Consequently, these EOSs are composed of a hot core starting from a specified value of the crust-core transition density computed at zero $T$, supplemented by a cold crust contribution obtained from existing literature (see, e.g., \cite{bombaci1996,lu2019} and references therein). In the left panel of Fig.~\ref{fig: EOS WITH COLD + hot CRUST}, we illustrate this type of NS EOSs using the BCPM density functional at various temperatures while considering a cold crust. In this case, we have maintained the zero-temperature density transition point between the inner crust and the core at $n\,(T=0)=0.08$ fm$^{-3}$, as reported in \cite{sharma2015}. For temperatures $T\lesssim2$\,MeV, the discontinuity between the cold crust and the hot core EOS is acceptable. However, as the temperature increases, e.g., $T=5$, $10$, and $20$\,MeV, the discontinuity becomes more and more pronounced.

\begin{figure}
    \centering
{\includegraphics[height=0.25\textheight]{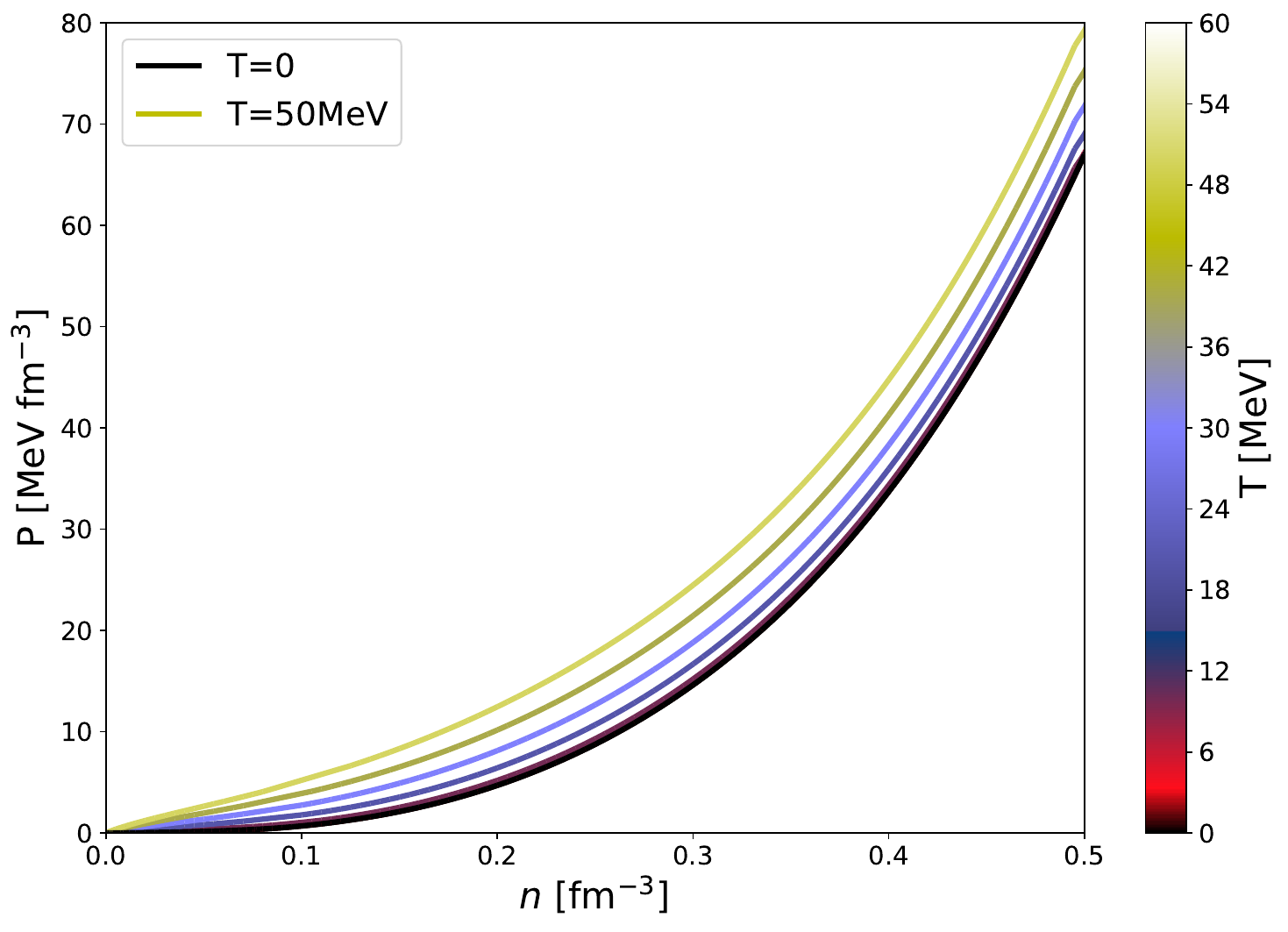}}
\caption{EOSs of $\beta$-stable neutrino-free matter at different temperature values ranging from $T=0$ to $50$\,MeV. 
For $T$ lower than $T_\mathrm{lim}=7.21$ MeV, we display the NS EOS containing the hot inner crust using the BCPM functional. At $T>7.21$\,MeV, the NS is composed of homogeneous dense matter, thus we illustrate the results of the bulk part of the hot BCPM functional.}
\label{fig: EOS hot crust}
\end{figure}

\begin{figure}
    \centering
{\includegraphics[height=0.25\textheight]{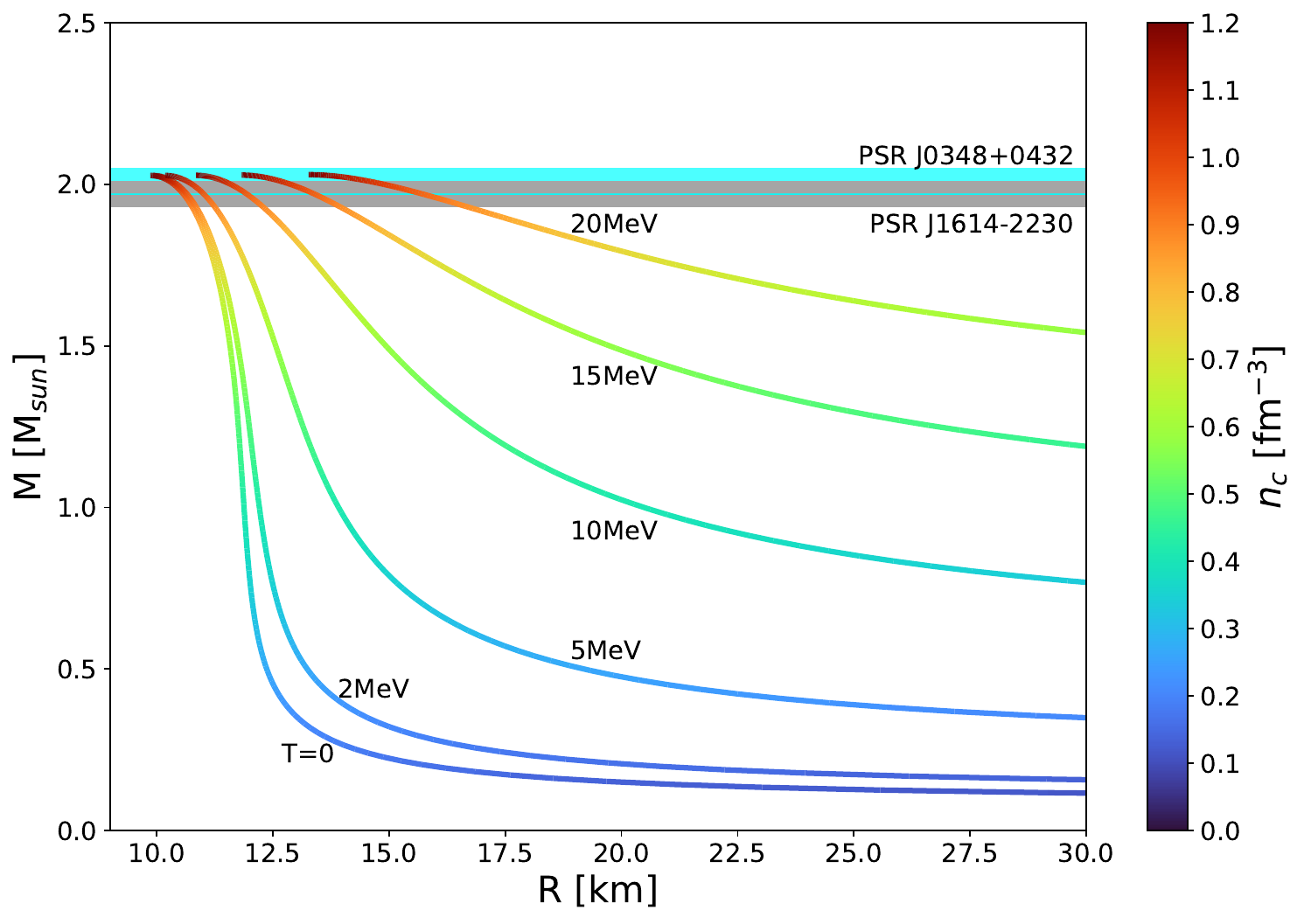}}
\caption{Computed mass--radius relations of NSs with an isothermal profile at temperatures of $T=0$, $2$, $5$, $15$, and $20$\,MeV. The observational mass constraints are $1.97 \pm 0.04$\,M$_\text{sun}$ (gray band) from the PSR J1614-2230 pulsar \citep{demorest2010} and $2.01 \pm 0.04$\,M$_\text{sun}$ (cyan band) from the PSR J0348+0432 pulsar \citep{antoniadis2013}. 
}
\label{fig: MvsR}
\end{figure}

When a constant $T$ is considered (isothermal profile), the radius of the star cannot be defined by the usual condition of vanishing pressure at the surface of the star. This is because at finite $T$, the pressure does not vanish even at very low densities \citep{buchler1977}, and, consequently, the thermal effect causes isolated NSs to expand. To address this, a temperature drop from the hot interior to the surface of the star is necessary. 
A proper determination of the temperature profile of the star would require dynamical simulations of thermal transport coupled with the $T$-dependent EOS. Given the difficulty and existing uncertainties,
other alternatives have been explored in the literature, such as using a neutrino sphere to ensure the temperature drops to zero at low density \citep{gondek1997}. For this reason, in our subsequent calculations for modeling NSs, we coupled the finite-temperature EOS of the core and the crust at $n = 10^{-4}$\,fm$^{-3}$ with the cold outer crust to obtain a cold surface.
The rationale for selecting the value of $10^{-4}$\,fm$^{-3}$ aligns with the outer-to-inner crust transition density of the BCPM functional at $T=0$ \citep{sharma2015}. Moreover, this value is compatible with the range of densities predicted in the literature where the neutrino sphere is employed \citep{gondek1997,strobel1999,fischer2009}. Therefore, our EOS at finite $T$ for NS calculations is built up as follows. For temperatures below the limiting temperature of $7.21$\,MeV, we compute the EOS taking into account the hot inner crust and the uniform matter according to the different density regions shown in Fig.~\ref{fig: transition density}, Fig.~\ref{fig:FoAdiff} and Table~\ref{tab: transition density}. When the temperature exceeds $7.21$\,MeV, the existence of the hot inner crust ceases; consequently, we employ calculations for hot uniform matter throughout. For all $T$ values, we rely on the cold outer crust for densities below $10^{-4}$\,fm$^{-3}$. 
It is worth mentioning that we have compared our NS mass-radius (M--R) relation results for $T=15$\,MeV (see Fig.~\ref{fig: MvsR}) with those from \cite{gondek1997}, which were obtained by considering an isothermal temperature profile at \(T=15\) MeV and a neutrino sphere, instead of a cold outer crust. We found that our results are consistent with theirs.

Plots of our EOSs at different temperatures with the hot inner crust prescription are depicted in the right panel of Fig.~\ref{fig: EOS WITH COLD + hot CRUST}. The impact of the hot temperature EOS in the crustal region can be appreciated at baryon densities below $0.08$\,fm$^{-3}$ by comparing the left and right panels of this figure. In Fig.~\ref{fig: EOS hot crust}, we plot a wider range of the EOSs at finite temperatures, extending up to $T=50$\,MeV. A consistent trend of growth is observed for all temperatures.

The M--R relations of NSs composed of neutrino-free $\beta$-stable matter
are illustrated in Fig.~\ref{fig: MvsR} at different temperature values.
The colorbar reflects the central density of the star. The M--R relations have been calculated by solving the Tolman-Oppenheimer-Volkoff equations \citep{shapiro1983}.
The observational masses used to depict the constraint of $M \sim 2\,M_{\text{sun}}$ in the M--R graph are $1.97 \pm 0.04$\,$M_{\text{sun}}$ from the PSR J1614-2230 pulsar \citep{demorest2010} and $2.01 \pm 0.04$\,$M_{\text{sun}}$ from the PSR J0348+0432 pulsar \citep{antoniadis2013}. 
The $T=0$ EOS satisfies this limit for the maximum mass, as well as the 1.4\,$M_{\text{sun}}$ radius constraint of $R_{1.4}= 11.9\pm1.4$\,km extracted by the LIGO-Virgo collaboration from the analysis of GW170817 \citep{ligovirgo2018}, and
is also compatible with the mass and radius constraints of $2.08\pm0.07$\,$M_{\text{sun}}$ measured in the PSR J0740+6620 pulsar \citep{fonseca2021} and $R_{1.4}= 12.45\pm0.65$\,km 
deduced from NICER data on PSR J0740+6620 \citep{miller2021}. We see in Fig.~\ref{fig: MvsR} that
the maximum mass of a NS slightly increases as $T$ grows, compared to the cold $T = 0$ case, but remains almost constant with temperature, in consonance
with previous findings \citep{bombaci1996,burgio2010,lu2019}. 
As $T$ increases, there is, however, a strong effect on the stellar radius, 
which for the same mass shifts towards a larger value when the star is hot.
As a consequence the compactness of the NS, $\mathcal{C}=G M/(R c^2)$, decreases with higher temperatures. For instance, if we consider the maximum mass configurations, we have $\mathcal{C}=0.30$ for $T=0$, whereas $\mathcal{C}=0.23$ for $T=20$\,MeV. This temperature dependence of the compactness may have observable consequences in the simulations of core-collapse supernovae, PNSs, and hot remnants of binary NS mergers \citep{oechslin2007}. On the other hand, as evident from the colorbar, for a given $T$ the central density takes the highest value for the maximum mass and decreases as the NS mass decreases and the NS radius increases.
Furthermore, for a given value of the mass, specially for not too small masses, the central density of the star is rather independent of temperature, showing a mild reduction with higher $T$. For example, for $M = 1.5 M_{\text{sun}}$ ($2 M_{\text{sun}}$) the central density experiences a slight decrease of 5\% (2.5\%) from $T=0$ to $T=20$\,MeV.

\begin{figure}
    \centering
{\includegraphics[height=0.25\textheight]{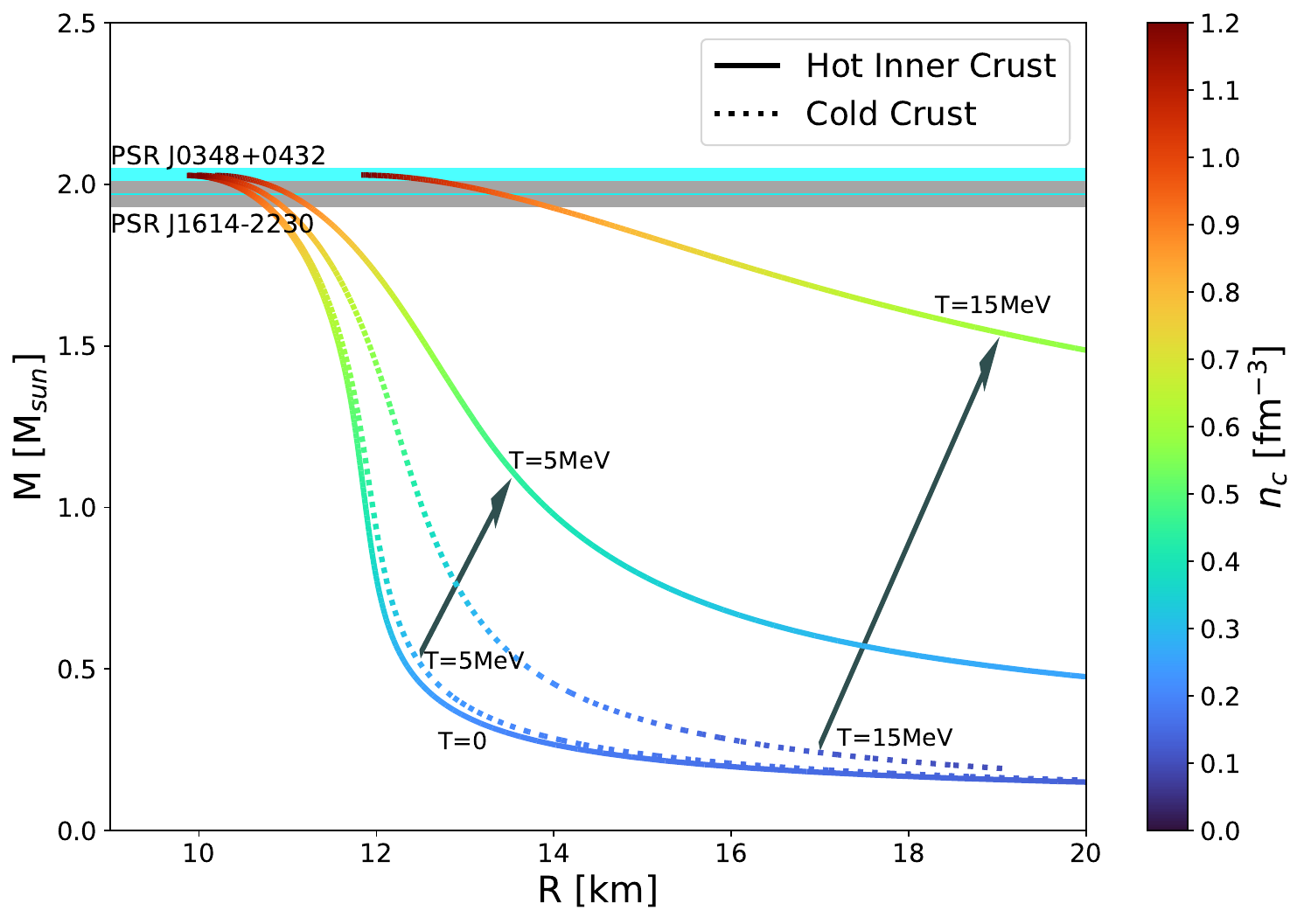}}
\caption{Mass--radius relations of NSs with an isothermal profile at temperatures of $T=0$, $5$, and $15$\,MeV. The dotted lines correspond to EOSs with cold crust and the solid lines correspond to EOSs with hot crust (the same as in Fig.~\ref{fig: MvsR}). Note that the horizontal scale goes up to $20$~km, whereas in Fig.~\ref{fig: MvsR} it goes up to $30$\,km. The observational constrains are the same as in Fig.~\ref{fig: MvsR}. The two arrows indicate the shift in the M--R relation for $T=5$ and $T=15$\,MeV.}
\label{fig: MvsR coldcrust}
\end{figure}

To emphasize the impact of the hot inner crust, we present a comparison of the M--R relations in Fig.~\ref{fig: MvsR coldcrust}. On the one hand, we consider a full cold crust below a density of $0.08$\,fm$^{-3}$ (dotted lines), and on the other hand, a hot inner crust (solid lines, the same of Fig.~\ref{fig: MvsR}) at $T=5$ and $15$\,MeV. We include the M--R relation at $T=0$ as a reference. The two arrows indicate the shift in the M--R relation for $T=5$ and $T=15$\,MeV when considering the hot inner crust. The EOSs with a cold crust predict M--R relations that are similar to those at $T=0$, particularly for low temperature values, e.g., $T=5$\,MeV, and around the maximum mass, i.e., $M\sim 2$\,M$_\text{sun}$. 
However, with the hot inner crust, the influence of temperature on the M--R relation becomes much more pronounced.
Clearly, when one attaches a cold crust below the cold crust-core transition density ($0.08$\,fm$^{-3}$) to the hot core, the  effect of temperature on the radius of the star is largely underestimated.
That is, even if a hot core EOS is implemented, with a cold crust the NS radius increases very moderately compared with the $T=0$ EOS.
When comparing the central densities predicted by considering a full cold crust and a hot inner crust (refer to Fig.~\ref{fig: MvsR coldcrust}), we notice that, for a given NS radius, the central density shifts to larger values when the hot crust is considered.

\section{Conclusion}
\label{sec: conclusion}

In \cite{sharma2015}, the BCPM energy density functional was used to derive a unified EOS for cold NSs based on ab-initio BHF calculations. 
Astrophysical studies of core-collapse supernovae, formation of PNSs, or the remnants of binary NS mergers, furthermore demand modeling hot compact objects that reach high temperatures.
Here, we predicted the thermal properties of stellar matter by generalizing at finite temperature the study with the BCPM functional.
To ascertain the EOS of the hot crust across various temperature values, we conducted hot self-consistent TF calculations using the BCPM functional involving spherical droplet configurations within WS cells.
In this first exploration of a unified hot EOS with the BCPM functional, we were largely driven by investigating the temperature effects on the baryonic matter and developed our study for isothermal $\beta$-equilibrated, neutrino-transparent configurations.

A prominent feature of our findings is that in hot stars there is a strong reduction of the value of the crust-core transition density.
The microscopic calculations predict that the inner crustal region completely dissolves at temperatures exceeding $7.21$\,MeV, a point we referred to as the limiting temperature. Moreover,
our analysis of the hot inner crust has unveiled two distinct branches of the interface between uniform and clustered matter below the limiting temperature. The branch with higher density values represents the upper transition density, marking the transition between the inner crust and the core; similar to the crust-core transition at zero temperature. Conversely, the branch with lower density values indicates a lower transition density, which does not occur at zero temperature, suggesting the reappearance of uniform matter in the outer layers of the hot NS.

Furthermore, we conducted a study on the M--R relation at finite temperature.
Here, the hot EOS obtained in our calculations was complemented by the cold BCPM EOS for the outer crust \citep{sharma2015}, which extends up to a density of $10^{-4}$\,fm$^{-3}$. This step was taken to ensure vanishing pressure at the surface of the star. A more accurate, and desirable, exploration needs a temperature profile of the star, but the latter is still uncertain; hence, we left the formulation of the calculations under a NS temperature profile for a future study.
Due to the existence of the limiting temperature, above $T_\mathrm{lim}=7.21$\,MeV the star consists entirely of uniform matter at densities greater than those corresponding to the border of the cold outer crust ($10^{-4}$\,fm$^{-3}$).
Our findings highlight the growing relevance of thermal effects on the M--R relation. On the one hand, for a given NS mass, the radius increases with temperature, and this effect is more pronounced at higher temperatures and at low stellar masses. On the other hand, for a given NS radius, the mass also increases with temperature but saturates to a value close to the maximum mass of a cold NS for sufficiently high temperatures.
When comparing our results with those obtained with a cold crust attached to the hot core (see Fig.~\ref{fig: MvsR coldcrust}), we observed
that the thermal effects from the hot core on the M--R relation are moderate, while the influence of the hot crust on the M--R relation is much more substantial.

To summarize, in this study, we successfully extended the BCPM EOS, for neutrino-free $\beta$-stable matter, to finite temperatures, thereby ensuring a meticulous treatment of the inner crust. The BCPM energy density functional offers a unified microscopic EOS for NSs under both cold and hot conditions. Furthermore, we determined M--R relations and central densities for hot compact objects at various temperature levels in an isothermal picture. Our results suggest that for NSs in hot environments, such as the end products of NSs merger events, it is essential to incorporate the hot inner crust.

\begin{acknowledgements}
This investigation was initiated in collaboration with Prof.~Artur Polls ($\dagger$\:12 August 2020), who made significant contributions to this study. We would like to dedicate this work in honor of his memory. We acknowledge useful discussions with J.~N.~De and J.~A.~Pons. CD is supported by the Prometeo Excellence Programme grant CIPROM/2022/13 from the Generalitat Valenciana, ERC Consolidator Grant “MAGNESIA” No. 817661, and has partial support from NORDITA and grant SGR2021-01269 while working on this article. This work is also partially supported by the María de Maeztu CEX2020-001058-M Excellence Unit Programme. MC and XV acknowledge partial support from Grants No. PID2020-118758GB-I00 and No. CEX2019–000918-M (through the “Unit of Excellence María de Maeztu 2020-2023” award to ICCUB) from the Spanish MCIN/AEI/10.13039/501100011033.
\end{acknowledgements}

\bibliographystyle{aa} % style aa.bst
\bibliography{aanda}

\providecommand{\noopsort}[1]{}\providecommand{\singleletter}[1]{#1}%
\begin{thebibliography}{78}
\expandafter\ifx\csname natexlab\endcsname\relax\def\natexlab#1{#1}\fi

\bibitem[{{Abbott et al. (LIGO \& Virgo Collaboration)}(2017)}]{ligovirgo2017}
{Abbott et al. (LIGO \& Virgo Collaboration)}. 2017, Phys. Rev. Lett., 119,
  161101

\bibitem[{{Abbott et al. (LIGO \& Virgo Collaboration)}(2018)}]{ligovirgo2018}
{Abbott et al. (LIGO \& Virgo Collaboration)}. 2018, Phys. Rev. Lett., 121,
  161101

\bibitem[{{Aguilera} {et~al.}(2008){Aguilera}, {Pons}, \&
  {Miralles}}]{aguilera2008}
{Aguilera}, D.~N., {Pons}, J.~A., \& {Miralles}, J.~A. 2008, \aap, 486, 255

\bibitem[{{Antoniadis} {et~al.}(2013){Antoniadis}, {Freire}, {Wex}, {Tauris},
  {Lynch}, {van Kerkwijk}, {Kramer}, {Bassa}, {Dhillon}, {Driebe}, {Hessels},
  {Kaspi}, {Kondratiev}, {Langer}, {Marsh}, {McLaughlin}, {Pennucci}, {Ransom},
  {Stairs}, {van Leeuwen}, {Verbiest}, \& {Whelan}}]{antoniadis2013}
{Antoniadis}, J., {Freire}, P. C.~C., {Wex}, N., {et~al.} 2013, Science, 340,
  448

\bibitem[{{Audi} {et~al.}(2012){Audi}, {M.}, {A.~H.}, {F.~G.}, {MacCormick},
  {Xu}, \& {Pfeiffer}}]{audi2012}
{Audi}, G., {M.}, W., {A.~H.}, W., {et~al.} 2012, Chinese Physics C, 36, 002

\bibitem[{{Baiotti}(2019)}]{baiotti2019}
{Baiotti}, L. 2019, Progress in Particle and Nuclear Physics, 109, 103714

\bibitem[{Baldo(1999)}]{bookbaldo1999}
Baldo, M., ed. 1999, Nuclear Methods and The Nuclear Equation of State
  (Singapore: World Scientific)

\bibitem[{Baldo \& Ferreira(1999)}]{baldo1999}
Baldo, M. \& Ferreira, L.~S. 1999, Phys. Rev. C, 59, 682

\bibitem[{{Baldo} {et~al.}(2010){Baldo}, {Robledo}, {Schuck}, \&
  {Vi{\~n}as}}]{baldo2010}
{Baldo}, M., {Robledo}, L., {Schuck}, P., \& {Vi{\~n}as}, X. 2010, Journal of
  Physics G Nuclear Physics, 37, 064015

\bibitem[{Baldo {et~al.}(2013)Baldo, Robledo, Schuck, \& Vi\~nas}]{baldo2013}
Baldo, M., Robledo, L.~M., Schuck, P., \& Vi\~nas, X. 2013, Phys. Rev. C, 87,
  064305

\bibitem[{{Baldo} {et~al.}(2017){Baldo}, {Robledo}, {Schuck}, \&
  {Vi{\~n}as}}]{baldo2017}
{Baldo}, M., {Robledo}, L.~M., {Schuck}, P., \& {Vi{\~n}as}, X. 2017, \prc, 95,
  014318

\bibitem[{{Baldo} {et~al.}(2008){Baldo}, {Schuck}, \& {Vi{\~n}as}}]{baldo2008}
{Baldo}, M., {Schuck}, P., \& {Vi{\~n}as}, X. 2008, Physics Letters B, 663, 390

\bibitem[{Barranco \& Buchler(1981)}]{barranco1981}
Barranco, M. \& Buchler, J.-R. 1981, Phys. Rev. C, 24, 1191

\bibitem[{{Barr{\`e}re} {et~al.}(2022){Barr{\`e}re}, {Guilet}, {Reboul-Salze},
  {Raynaud}, \& {Janka}}]{barrere2022}
{Barr{\`e}re}, P., {Guilet}, J., {Reboul-Salze}, A., {Raynaud}, R., \& {Janka},
  H.~T. 2022, \aap, 668, A79

\bibitem[{Bauswein {et~al.}(2013)Bauswein, Baumgarte, \& Janka}]{bauswein2013}
Bauswein, A., Baumgarte, T.~W., \& Janka, H.-T. 2013, Phys. Rev. Lett., 111,
  131101

\bibitem[{{Baym} {et~al.}(1971a){Baym}, {Bethe}, \& {Pethick}}]{baym1971a}
{Baym}, G., {Bethe}, H.~A., \& {Pethick}, C.~J. 1971a, Nucl. Phys. A, 175, 225

\bibitem[{{Baym} {et~al.}(1971){Baym}, {Pethick}, \& {Sutherland}}]{baym1971b}
{Baym}, G., {Pethick}, C., \& {Sutherland}, P. 1971, \apj, 170, 299

\bibitem[{{Beloborodov} \& {Li}(2016)}]{beloborodov16}
{Beloborodov}, A.~M. \& {Li}, X. 2016, \apj, 833, 261

\bibitem[{Bethe(1990)}]{bethe1990}
Bethe, H.~A. 1990, Rev. Mod. Phys., 62, 801

\bibitem[{{Bombaci}(1996)}]{bombaci1996}
{Bombaci}, I. 1996, \aap, 305, 871

\bibitem[{Brack {et~al.}(1985)Brack, Guet, \& Håkansson}]{brack1985}
Brack, M., Guet, C., \& Håkansson, H.-B. 1985, Physics Reports, 123, 275

\bibitem[{{Buchler} \& {Coon}(1977)}]{buchler1977}
{Buchler}, J.~R. \& {Coon}, S.~A. 1977, \apj, 212, 807

\bibitem[{{Burgio} \& {Fantina}(2018)}]{burgio2018}
{Burgio}, F. \& {Fantina}, A.~F. 2018, in Astrophysics and Space Science
  Library, Vol. 457, Astrophysics and Space Science Library, ed. L.~{Rezzolla},
  P.~{Pizzochero}, D.~I. {Jones}, N.~{Rea}, \& I.~{Vida{\~n}a} (Springer), 255

\bibitem[{{Burgio} {et~al.}(2007){Burgio}, {Baldo}, {Nicotra}, \&
  {Schulze}}]{burgio2007}
{Burgio}, G.~F., {Baldo}, M., {Nicotra}, O.~E., \& {Schulze}, H.~J. 2007,
  \apss, 308, 387

\bibitem[{{Burgio} \& {Schulze}(2010)}]{burgio2010}
{Burgio}, G.~F. \& {Schulze}, H.~J. 2010, \aap, 518, A17

\bibitem[{{Coti Zelati} {et~al.}(2018){Coti Zelati}, {Rea}, {Pons}, {Campana},
  \& {Esposito}}]{cotizelati18}
{Coti Zelati}, F., {Rea}, N., {Pons}, J.~A., {Campana}, S., \& {Esposito}, P.
  2018, \mnras, 474, 961

\bibitem[{Dehman {et~al.}(2023)Dehman, Pons, Viganò, \& Rea}]{dehman2023}
Dehman, C., Pons, J.~A., Viganò, D., \& Rea, N. 2023, \mnras, 520, L42

\bibitem[{{Dehman} {et~al.}(2020){Dehman}, {Vigan{\`o}}, {Rea}, {Pons},
  {Perna}, \& {Garcia-Garcia}}]{dehman2020}
{Dehman}, C., {Vigan{\`o}}, D., {Rea}, N., {et~al.} 2020, \apjl, 902, L32

\bibitem[{{Demorest} {et~al.}(2010){Demorest}, {Pennucci}, {Ransom}, {Roberts},
  \& {Hessels}}]{demorest2010}
{Demorest}, P.~B., {Pennucci}, T., {Ransom}, S.~M., {Roberts}, M.~S.~E., \&
  {Hessels}, J.~W.~T. 2010, Nature, 467, 1081

\bibitem[{{Figura} {et~al.}(2021){Figura}, {Li}, {Lu}, {Burgio}, {Li}, \&
  {Schulze}}]{figura2021}
{Figura}, A., {Li}, F., {Lu}, J.-J., {et~al.} 2021, \prd, 103, 083012

\bibitem[{{Fischer} {et~al.}(2009){Fischer}, {Whitehouse}, {Mezzacappa},
  {Thielemann}, \& {Liebendörfer}}]{fischer2009}
{Fischer}, T., {Whitehouse}, S.~C., {Mezzacappa}, A., {Thielemann}, F.-K., \&
  {Liebendörfer}, M. 2009, A\&A, 499, 1

\bibitem[{{Flanagan} \& {Hinderer}(2008)}]{flanagan2008}
{Flanagan}, {\'E}.~{\'E}. \& {Hinderer}, T. 2008, \prd, 77, 021502

\bibitem[{{Fonseca} {et~al.}(2021){Fonseca}, {Cromartie}, {Pennucci}, {Ray},
  {Kirichenko}, {Ransom}, {Demorest}, {Stairs}, {Arzoumanian}, {Guillemot},
  {Parthasarathy}, {Kerr}, {Cognard}, {Baker}, {Blumer}, {Brook}, {DeCesar},
  {Dolch}, {Dong}, {Ferrara}, {Fiore}, {Garver-Daniels}, {Good}, {Jennings},
  {Jones}, {Kaspi}, {Lam}, {Lorimer}, {Luo}, {McEwen}, {McKee}, {McLaughlin},
  {McMann}, {Meyers}, {Naidu}, {Ng}, {Nice}, {Pol}, {Radovan},
  {Shapiro-Albert}, {Tan}, {Tendulkar}, {Swiggum}, {Wahl}, \&
  {Zhu}}]{fonseca2021}
{Fonseca}, E., {Cromartie}, H.~T., {Pennucci}, T.~T., {et~al.} 2021, \apjl,
  915, L12

\bibitem[{{Gondek} {et~al.}(1997){Gondek}, {Haensel}, \& {Zdunik}}]{gondek1997}
{Gondek}, D., {Haensel}, P., \& {Zdunik}, J.~L. 1997, \aap, 325, 217

\bibitem[{{Gonzalez-Boquera} {et~al.}(2019){Gonzalez-Boquera}, {Centelles},
  {Vi{\~n}as}, \& {Routray}}]{gonzalez2019}
{Gonzalez-Boquera}, C., {Centelles}, M., {Vi{\~n}as}, X., \& {Routray}, T.~R.
  2019, \prc, 100, 015806

\bibitem[{Haensel {et~al.}(2007)Haensel, Potekhin, \& Yakovlev}]{haensel2007}
Haensel, P., Potekhin, A.~Y., \& Yakovlev, D.~G. 2007, {Neutron stars 1:
  Equation of state and structure}, Vol. 326 (New York, USA: Springer)

\bibitem[{{Hempel} \& {Schaffner-Bielich}(2010)}]{hempel2010}
{Hempel}, M. \& {Schaffner-Bielich}, J. 2010, \nphysa, 837, 210

\bibitem[{{Hinderer} {et~al.}(2010){Hinderer}, {Lackey}, {Lang}, \&
  {Read}}]{hinderer2010}
{Hinderer}, T., {Lackey}, B.~D., {Lang}, R.~N., \& {Read}, J.~S. 2010, \prd,
  81, 123016

\bibitem[{{Krastev} \& {Li}(2019)}]{krastev2019}
{Krastev}, P.~G. \& {Li}, B.-A. 2019, Journal of Physics G Nuclear Physics, 46,
  074001

\bibitem[{{Kubis}(2004)}]{kubis2004}
{Kubis}, S. 2004, \prc, 70, 065804

\bibitem[{{Kumar} \& {Bo{\v{s}}njak}(2020)}]{kumar2020}
{Kumar}, P. \& {Bo{\v{s}}njak}, {\v{Z}}. 2020, \mnras, 494, 2385

\bibitem[{{Lattimer} \& {Swesty}(1991)}]{lattimer1991}
{Lattimer}, J.~M. \& {Swesty}, D.~F. 1991, \nphysa, 535, 331

\bibitem[{{Liebend{\"o}rfer} {et~al.}(2005){Liebend{\"o}rfer}, {Rampp},
  {Janka}, \& {Mezzacappa}}]{liebendoerfer2005}
{Liebend{\"o}rfer}, M., {Rampp}, M., {Janka}, H.~T., \& {Mezzacappa}, A. 2005,
  \apj, 620, 840

\bibitem[{{Logoteta} {et~al.}(2021){Logoteta}, {Perego}, \&
  {Bombaci}}]{logoteta2021}
{Logoteta}, D., {Perego}, A., \& {Bombaci}, I. 2021, \aap, 646, A55

\bibitem[{{Lorenz} {et~al.}(1993){Lorenz}, {Ravenhall}, \&
  {Pethick}}]{lorenz1993}
{Lorenz}, C.~P., {Ravenhall}, D.~G., \& {Pethick}, C.~J. 1993, \prl, 70, 379

\bibitem[{{Lu} {et~al.}(2019){Lu}, {Li}, {Burgio}, {Figura}, \&
  {Schulze}}]{lu2019}
{Lu}, J.-J., {Li}, Z.-H., {Burgio}, G.~F., {Figura}, A., \& {Schulze}, H.~J.
  2019, \prc, 100, 054335

\bibitem[{Marques {et~al.}(2017)Marques, Oertel, Hempel, \&
  Novak}]{marques2017}
Marques, M., Oertel, M., Hempel, M., \& Novak, J. 2017, Phys. Rev. C, 96,
  045806

\bibitem[{{Menezes} \& {Provid{\^e}ncia}(2017)}]{menezes2017}
{Menezes}, D.~P. \& {Provid{\^e}ncia}, C. 2017, \prc, 96, 045803

\bibitem[{{Miller} {et~al.}(2021){Miller}, {Lamb}, {Dittmann}, {Bogdanov},
  {Arzoumanian}, {Gendreau}, {Guillot}, {Ho}, {Lattimer}, {Loewenstein},
  {Morsink}, {Ray}, {Wolff}, {Baker}, {Cazeau}, {Manthripragada}, {Markwardt},
  {Okajima}, {Pollard}, {Cognard}, {Cromartie}, {Fonseca}, {Guillemot}, {Kerr},
  {Parthasarathy}, {Pennucci}, {Ransom}, \& {Stairs}}]{miller2021}
{Miller}, M.~C., {Lamb}, F.~K., {Dittmann}, A.~J., {et~al.} 2021, \apjl, 918,
  L28

\bibitem[{{Moustakidis}(2012)}]{moustakidis2012}
{Moustakidis}, C.~C. 2012, \prc, 86, 015801

\bibitem[{{Neill} {et~al.}(2023){Neill}, {Preston}, {Newton}, \&
  {Tsang}}]{neill2023}
{Neill}, D., {Preston}, R., {Newton}, W.~G., \& {Tsang}, D. 2023, \prl, 130,
  112701

\bibitem[{{Oechslin} {et~al.}(2007){Oechslin}, {Janka}, \&
  {Marek}}]{oechslin2007}
{Oechslin}, R., {Janka}, H.~T., \& {Marek}, A. 2007, \aap, 467, 395

\bibitem[{{Oertel} {et~al.}(2017){Oertel}, {Hempel}, {Kl{\"a}hn}, \&
  {Typel}}]{oertel2017}
{Oertel}, M., {Hempel}, M., {Kl{\"a}hn}, T., \& {Typel}, S. 2017, Reviews of
  Modern Physics, 89, 015007

\bibitem[{Pearson \& Chamel(2022)}]{pearson2022}
Pearson, J.~M. \& Chamel, N. 2022, Phys. Rev. C, 105, 015803

\bibitem[{Pearson {et~al.}(2020)Pearson, Chamel, \& Potekhin}]{pearson2020}
Pearson, J.~M., Chamel, N., \& Potekhin, A.~Y. 2020, \prc, 101, 015802

\bibitem[{{Pearson} {et~al.}(2018){Pearson}, {Chamel}, {Potekhin}, {Fantina},
  {Ducoin}, {Dutta}, \& {Goriely}}]{pearson2018}
{Pearson}, J.~M., {Chamel}, N., {Potekhin}, A.~Y., {et~al.} 2018, \mnras, 481,
  2994

\bibitem[{{Pi} {et~al.}(1986){Pi}, {Vinas}, {Barranco}, {Polls}, \&
  {Perez-Canyellas}}]{pi1986}
{Pi}, M., {Vinas}, X., {Barranco}, M., {Polls}, A., \& {Perez-Canyellas}, A.
  1986, \aaps, 64, 439

\bibitem[{Piekarewicz {et~al.}(2014)Piekarewicz, Fattoyev, \&
  Horowitz}]{piekarewicz2014}
Piekarewicz, J., Fattoyev, F.~J., \& Horowitz, C.~J. 2014, Phys. Rev. C, 90,
  015803

\bibitem[{{Pons} {et~al.}(1999){Pons}, {Reddy}, {Prakash}, {Lattimer}, \&
  {Miralles}}]{pons1999}
{Pons}, J.~A., {Reddy}, S., {Prakash}, M., {Lattimer}, J.~M., \& {Miralles},
  J.~A. 1999, \apj, 513, 780

\bibitem[{{Pons} \& {Vigan{\`o}}(2019)}]{pons2019}
{Pons}, J.~A. \& {Vigan{\`o}}, D. 2019, Living Reviews in Computational
  Astrophysics, 5, 3

\bibitem[{{Pooley} {et~al.}(2018){Pooley}, {Kumar}, {Wheeler}, \&
  {Grossan}}]{pooley2018}
{Pooley}, D., {Kumar}, P., {Wheeler}, J.~C., \& {Grossan}, B. 2018, \apjl, 859,
  L23

\bibitem[{{Potekhin} {et~al.}(2015){Potekhin}, {Pons}, \&
  {Page}}]{potekhin2015}
{Potekhin}, A.~Y., {Pons}, J.~A., \& {Page}, D. 2015, \ssr, 191, 239

\bibitem[{{Raithel} {et~al.}(2021){Raithel}, {Paschalidis}, \&
  {{\"O}zel}}]{raithel2021}
{Raithel}, C.~A., {Paschalidis}, V., \& {{\"O}zel}, F. 2021, \prd, 104, 063016

\bibitem[{{Reddy} {et~al.}(1999){Reddy}, {Prakash}, {Lattimer}, \&
  {Pons}}]{reddy1999}
{Reddy}, S., {Prakash}, M., {Lattimer}, J.~M., \& {Pons}, J.~A. 1999, \prc, 59,
  2888

\bibitem[{Sekiguchi {et~al.}(2011)Sekiguchi, Kiuchi, Kyutoku, \&
  Shibata}]{sekiguchi2011}
Sekiguchi, Y., Kiuchi, K., Kyutoku, K., \& Shibata, M. 2011, Phys. Rev. Lett.,
  107, 051102

\bibitem[{{Shapiro} \& {Teukolsky}(1983)}]{shapiro1983}
{Shapiro}, S.~L. \& {Teukolsky}, S.~A. 1983, {Black holes, white dwarfs and
  neutron stars. The physics of compact objects} (Wiler)

\bibitem[{{Sharma} {et~al.}(2015){Sharma}, {Centelles}, {Vi{\~n}as}, {Baldo},
  \& {Burgio}}]{sharma2015}
{Sharma}, B.~K., {Centelles}, M., {Vi{\~n}as}, X., {Baldo}, M., \& {Burgio},
  G.~F. 2015, \aap, 584, A103

\bibitem[{Shibata(2015)}]{shibata2015}
Shibata, M. 2015, Numerical relativity, Vol.~1 (World Scientific)

\bibitem[{{Sil} {et~al.}(2002){Sil}, {De}, {Samaddar}, {Vi{\~n}as},
  {Centelles}, {Agrawal}, \& {Patra}}]{sil2002}
{Sil}, T., {De}, J.~N., {Samaddar}, S.~K., {et~al.} 2002, \prc, 66, 045803

\bibitem[{{Soma} \& {Bandyopadhyay}(2020)}]{soma2020}
{Soma}, S. \& {Bandyopadhyay}, D. 2020, \apj, 890, 139

\bibitem[{{Sotani} {et~al.}(2012){Sotani}, {Nakazato}, {Iida}, \&
  {Oyamatsu}}]{sotani2012}
{Sotani}, H., {Nakazato}, K., {Iida}, K., \& {Oyamatsu}, K. 2012, \prl, 108,
  201101

\bibitem[{{Steiner} \& {Watts}(2009)}]{steiner2009}
{Steiner}, A.~W. \& {Watts}, A.~L. 2009, \prl, 103, 181101

\bibitem[{{Strobel} {et~al.}(1999){Strobel}, {Schaab}, \&
  {Weigel}}]{strobel1999}
{Strobel}, K., {Schaab}, C., \& {Weigel}, M.~K. 1999, \aap, 350, 497

\bibitem[{{Suraud}(1987)}]{suraud1987}
{Suraud}, E. 1987, \nphysa, 462, 109

\bibitem[{Taranto {et~al.}(2013)Taranto, Baldo, \& Burgio}]{taranto2013}
Taranto, G., Baldo, M., \& Burgio, G.~F. 2013, Phys. Rev. C, 87, 045803

\bibitem[{Thi {et~al.}(2021)Thi, Mondal, \& Gulminelli}]{thi2021}
Thi, H.~D., Mondal, C., \& Gulminelli, F. 2021, Universe, 7, 373

\bibitem[{Wiringa {et~al.}(1995)Wiringa, Stoks, \& Schiavilla}]{wiringa1995}
Wiringa, R.~B., Stoks, V. G.~J., \& Schiavilla, R. 1995, Phys. Rev. C, 51, 38

\bibitem[{{Xu} {et~al.}(2009){Xu}, {Chen}, {Li}, \& {Ma}}]{xu2009}
{Xu}, J., {Chen}, L.-W., {Li}, B.-A., \& {Ma}, H.-R. 2009, \apj, 697, 1549

\end{thebibliography}

\end{document}